\documentstyle[12pt,amssymb,amscd,righttag]{amsart}

\newtheorem{thm}{Theorem}[section]
\newtheorem{lem}{Lemma}[section]
\newtheorem{pro}{Proposition}[section]
\newtheorem{cor}{Corollary}[section]

\theoremstyle{definition}

\theoremstyle{remark}

\newtheorem{rem}{Remark}[section]

\numberwithin{equation}{section}

\renewcommand{\phi}{\varphi}

\newcommand{\ttheta}{\tilde{\theta}}

\renewcommand{\Re}{\operatorname{Re}}
\renewcommand{\Im}{\operatorname{Im}}
\newcommand{\Spec}{\operatorname{Spec}}
\newcommand{\Tr}{\operatorname{Tr}}
\newcommand{\TR}{\operatorname{TR}}
\newcommand{\tr}{\operatorname{tr}}
\newcommand{\Id}{\operatorname{Id}}
\newcommand{\ord}{\operatorname{ord}}
\newcommand{\SEll}{\operatorname{SEll}}
\newcommand{\Ell}{\operatorname{Ell}}
\newcommand{\End}{\operatorname{End}}
\newcommand{\Res}{\operatorname{Res}}
\newcommand{\res}{\operatorname{res}}
\newcommand{\Ad}{\operatorname{Ad}}
\newcommand{\Ker}{\operatorname{Ker}}
\newcommand{\rk}{\operatorname{rk}}
\newcommand{\Aut}{\operatorname{Aut}}
\newcommand{\Map}{\operatorname{Map}}
\newcommand{\Mon}{\operatorname{Mon}}
\newcommand{\mon}{\operatorname{mon}}
\newcommand{\Diff}{\operatorname{Diff}}
\newcommand{\Image}{\operatorname{Image}}

\newcommand{\wC}{\Bbb C}
\newcommand{\wR}{\Bbb R}
\newcommand{\wZ}{\Bbb Z}
\newcommand{\wG}{\bold G}

\newcommand{\df}{\partial}
\newcommand{\zuo}{\wZ_+\cup 0}
\newcommand{\sG}{\widetilde{G}}
\newcommand{\sx}{\widetilde{x}}
\newcommand{\sT}{\widetilde{T}}

\newcommand{\lrs}{\widetilde{\phantom{abab}}}
\newcommand{\rs}{@>>\lrs>}
\newcommand{\frg}{\frak{g}}
\newcommand{\sfrg}{\widetilde{\frg}}
\newcommand{\frh}{\frak{h}}
\newcommand{\fell}{\frak{e}\frak{l}\frak{l}}
\newcommand{\cD}{\cal{D}}
\newcommand{\bs}{\bold{s}}

\begin{document}

\pagestyle{style}

\vspace{5mm}
\title{Geometry of determinants of elliptic operators}
\author{Maxim Kontsevich}
\address{Max-Planck-Institut f\"ur Mathematik, Gottfried-Claren-Stra{\ss}e
26, 53225 Bonn, Germany \\
Department of Mathematics, University of California,
Berkeley, CA 94720}
\email{maxim@@mpim-bonn.mpg.de}
\author{Simeon Vishik}
\address{Max-Planck-Institut f\"ur Mathematik, Gottfried-Claren-Stra{\ss}e
26, 53225 Bonn, Germany \\
Department of Mathematics, Temple University, Philadelphia, PA 19122}
\email{senia@@mpim-bonn.mpg.de}
\date{June 1994}
\maketitle

\tableofcontents
\section{Introduction}

D.B.~Ray and I.M.~Singer invented zeta-regularized determinants
for positive definite elliptic pseudo-differential operators (PDOs)
of positive
orders acting in the space of smooth sections of a finite-dimensional
vector bundle $E$ over a closed finite-dimensional manifold $M$
(\cite{RS1}, \cite{RS2}).

Recall that for any such invertible operator $A$ its zeta-function, defined
for $\Re s\gg 0$ by the formula
$$
\zeta_A(s)=\sum_{\{\lambda_i\}\in\Spec A}\lambda_i^{-s}=\Tr A^{-s}\,\,,
$$
has a meromorphic continuation to $\wC$ without pole at zero.
(Here, the sum includes the algebraic multiplicities.)
A zeta-regularized determinant of $A$ is, by definition,
$$
{\det}_\zeta(A):=\exp\left(-{\df\over\df s}\zeta_A(s)\big|_{s=0}\right)\,.
$$

We are interested in this paper in multiplicative properties of these
determinants, i.e., we want to compute the ratio
\begin{equation}
F(A,B):={\det}_\zeta(AB)/({\det}_\zeta(A){\det}_\zeta(B))\,\,\,.
\label{A40}
\end{equation}
We call it the multiplicative anomaly. In general, it is not equal to $1$.
For example for $A=\Delta+\Id$ and $B=\Delta+2\Id$, where $\Delta$
is the Laplacian acting on functions on an even-dimensional Riemannian
manifold, $F(A,B)$ is defined and it is almost never equal to $1$.

The determinant $\det_\zeta(A)$ is defined for an invertible elliptic PDO $A$,
$\ord A>0$, admitting a spectral cut. Such a cut exists, if $A$
satisfies the Agmon-Nirenberg condition formulated as follows
(for closed $M$). There exists a closed conical sector
$V=\left\{\lambda\colon\theta_1\le\arg\lambda\le\theta_2\right\}$,
$\theta_1<\theta_2$, in the spectral plane $\wC$ such that all eigenvalues
of the principal symbol $\sigma_d(A)(x,\xi)$ do not belong to $V$
for any $(x,\xi)\in T^*M\setminus M$. If such a condition is satisfied
for $A$, then in $V$ there is no more than a finite number of eigenvalues
of $A$ including their algebraic multiplicities.

Note that this condition on $\sigma_d(A)(x,\xi)$ cannot be satisfied
for any $d=\ord A\in\wC\setminus\wR$ because for any such $d$
the curve $\wR_+\ni t\to t^d\in\wC^{\times}$ crosses all the rays $L_\theta$
infinitely many times.
Note also that the Agmon-Nirenberg condition is formulated in terms
of the principal symbol of $A$.
So it is a micro-local condition, and it can be checked effectively.
It provides us with an information about the spectrum of $A$ which
we cannot compute in general.

Let us pick a cut $L_\theta=\left\{\lambda\colon\arg\lambda=\theta\right\}$,
$\theta_1<\theta<\theta_2$, such that $\Spec A\cap L_\theta=\emptyset$,
and define a zeta-function $\zeta_{A,\theta}(s)$ of $A$ corresponding
to this cut. Namely we define $A_{(\theta)}^{-s}$ for $\Re s\in\wR_+$ large
enough by
\begin{equation}
A_{(\theta)}^{-s}:={i\over 2\pi}\int_{\Gamma_{(\theta)}}\left(A-\lambda
\right)^{-1}\lambda_{(\theta)}^{-s}d\lambda\,\,\,,
\label{A15}
\end{equation}
where $\Gamma_{(\theta)}$ is a contour
$\Gamma_{1,\theta}(\rho)\cup\Gamma_{0,\theta}(\rho)\cup\Gamma_{2,\theta}
(\rho)$,
$\Gamma_{1,\theta}(\rho)\colon\!=\!\left\{\lambda\!=\!x\exp(i\theta),
+\!\infty\!>\!x\!\ge\!\rho\right\}$,
$\Gamma_{0,\theta}(\rho)\colon\!=\!\left\{\lambda\!=\!\rho\exp(i\phi),
\theta\!\ge\!\phi\!\ge\!\theta\!-\!2\pi\right\}$,
$\Gamma_{2,\theta}(\rho)\colon\!=\!\left\{\lambda\!=\!x\exp i(\theta\!-\!2
\pi),\rho\!\le\!x\!<\!+\!\infty\right\}$,
and $\rho$ is a positive number such that all the eigenvalues
in $\Spec(A)$ are outside of the disk
$D_\rho:=\{\lambda\colon |\lambda|\le\rho\}$. Here,
$\lambda_{(\theta)}^{-s}:=\exp\left(-s\log_{(\theta)}\lambda\right)$
with a branch $\log_{(\theta)}\lambda$,
$\theta\ge\Im\log_{(\theta)}\lambda\ge\theta-2\pi$.
Then a family $A_{(\theta)}^{-s}$ for any $s$ is defined
as $A^kA_{(\theta)}^{-(s+k)}$ for $k\in\wZ_+$ large enough (and depending
on $\Re s$). This definition of $A_{(\theta)}^{-s}$ is independent of $k$.
Then $\zeta_{A,\theta}(s)$ is defined as $\Tr\left(A_{(\theta)}^{-s}\right)$
for $\Re s$ large enough.%
\footnote{The trace $\Tr\left(A_{(\theta)}^{-s}\right)$
for $\Re(s\ord A)>\dim M$ is equal to the sum
$\sum\lambda_{i,(\theta)}^{-s}$ (including algebraic multiplicities)
as it follows from the Lidskii Theorem \cite{Li}, \cite{Re}, XI.}
This zeta-function has a meromorphic
continuation to the whole complex plane and it is regular at $s=0$.
This zeta-function depends on an admissible cut $L_\theta$.
Nevertheless the corresponding determinant is independent of such a cut
for $L_\theta\subset V$. The reason is that if the number $m$,
$m\in\zuo$, of eigenvalues of $A$ in the sector between $L_\theta$ and
$L_{\ttheta}$ is finite, then
$$
\df_s\left(\zeta_{A,\theta}(s)-\zeta_{A,\ttheta}(s)\right)\big|_{s=0}=\pm
2\pi im\,\,\,.
$$

Note that in general $\det_\zeta(A)$ depends on the homotopy class
of an admissible spectral sector $V$
 in the Agmon-Nirenberg condition for $A$.

The rest of the paper is devoted to the study of properties
of the multiplicative anomaly and related algebraic and geometric objects.
Using Fredholm determinants we introduce a central $\wC^{\times}$-extension
$\sG$ of the group
$G$ of elliptic symbols and a partially defined section $d_0$ of it.
All properties of multiplicative anomaly are encoded in these objects.

One of our results is an extension of the notion of the zeta-regularized
determinant to a larger class of operators (including operators
of nonreal orders). The modified definition of $\det(A)$ does not use
the existence of any holomorphic family $A^{-s}$ for a given $A$ and
does not use
any analytic continuations. The main tool is a new trace class
functional $\TR$ defined for classical PDOs of noninteger orders.
We discovered a simple Lie-algebraic description of $\sG$ and
of $d_0$ in a neighborhood of the identity $\Id\in\sG$ purely in terms
of symbols. There is an interesting interplay between invariant
quadratic forms and $2$-cocycles on Lie algebras.
We also describe an analogue of the determinant Lie group $\sG$
for a certain natural class of PDOs on odd-dimensional manifolds.
We prove, in particular, that for positive self-adjoint elliptic
differential operators on such manifolds the multiplicative property
holds.

This paper is essentially a compressed version of our previous paper
\cite{KV}. The aim of the current paper is to give a short and clear
exposition of our present understanding of the subject. In comparison
with \cite{KV} we change the general structure of the text and
present some new proofs. Here we drop minor details of the proofs
but try to give main ideas. Our notaitions differ a little from
the notations of \cite{KV}.

\subsection{Formula for multiplicative anomaly}

Let $A$ and $B$ be invertible elliptic PDOs of real nonzero orders
$\alpha$ and $\beta$ such that $\alpha+\beta\ne 0$ and such that
their principal symbols
$\sigma_\alpha(A)$, $\sigma_\beta(B)$, and $\sigma_{\alpha+\beta}(AB)$
 obey the Agmon-Nirenberg condition (with appropriate spectral
cuts). Let $A_t$ be a smooth deformation of the elliptic PDO $A=A_0$
such that $\ord A_t\equiv\ord A$. Hence $A_t$ and $A_tB$ satisfy
the Agmon-Nirenberg condition for small $t$. The complex powers
of $A_t$, $B$, and of $A_tB$ are defined for such $t$ by (\ref{A15})
with appropriate spectral cuts. Thus the determinants of these
operators are defined.

\begin{pro}
Under the conditions above, the variation formula for the multiplicative
anomaly (\ref{A40}) holds (for small $t$)
\begin{equation}
{\df\over\df t}\log F\left(A_t,B\right)=-\res\left(\sigma\left({\df\over\df t}
A_t\cdot A_t^{-1}\right)\cdot\sigma\left({\log \left(A_tB\right)\over\alpha+
\beta}-
{\log A\over\alpha}\right)\right).
\label{A46}
\end{equation}
\label{PA45}
\end{pro}

This formula is proved in \cite{KV}, Section~2.

The logarithms in (\ref{A46}) are defined as the derivatives at $s=0$
of complex
powers $\left(A_tB\right)^s$ and $B^s$. Note that
\;$\log\left(A_tB\right)/(\alpha+\beta)-\log A/\alpha\in CL^0$ (i.e.,
it is a classical PDO of order zero).

To remind, the noncommutative residue of a classical PDO-symbol $a$
of an integer order is equal to the integral over $M$, $\dim M=n$,
of a density defined by
\begin{equation}
\res_x(a):=\int_{S_x^*M}\tr\left(a_{-n}(x,\xi)\right)d'\xi\,\,\,.
\label{B7}
\end{equation}
This density on $M$ is independent of a choice of local coordinates
on $M$. The integral (\ref{B7}) is taken over the unit sphere
$S_x^*M=\left\{\xi\in T_x^*M\colon|\xi|=1\right\}$.

\begin{rem}
Using formula (\ref{A46}) one can obtain an explicit local expression
in terms of symbols for the multiplicative anomaly (\ref{A40}) if $A$
and $B$ are sufficiently close to positive definite self-adjoint PDOs.
Namely, in this case, one can connect $A$ with $A_1:=B^{\alpha/\beta}$
by a smooth path in the space of elliptic PDOs of order $\alpha$
admitting a spectral cut close to $\wR_-\subset\wC$.
\label{RA47}
\end{rem}

\begin{rem}
A formula for $F(A,B)$ for commuting self-adjoint positive elliptic DOs
was obtained by M.~Wodzicki, see \cite{Kas}. For noncommuting positive
self-adjoint elliptic PDOs a variation formula for $F(A,B)$ in a form
different from (\ref{A46}) was obtained by L.~Friedlander, \cite{Fr}.
\label{RA48}
\end{rem}

\section{Determinant Lie group}

%

{}From now on all elliptic PDOs are supposed to be invertible.
Let $A$, $B$, and $AB$ admit spectral cuts and let their orders
be nonzero real numbers. Then the multiplicative anomaly $F(A,B)$ depends
on symbols $\sigma(A)$ and $\sigma(B)$ only (for fixed admissible
spectral cuts). This statement is an immediate consequence
of the following lemma.

\begin{lem}
For an elliptic operator $A$, $\ord A\in\wR^{\times}$, admitting a spectral
cut $L_\theta$ and for any invertible operator $Q$ of the form $Q=\Id+S$,
where the Schwartz kernel of $S$ is $C^{\infty}$ on $M\times M$ (i.e.,
$S$ is smoothing), the equality holds
\begin{equation}
{\det}_\zeta(AQ)={\det}_\zeta(A){\det}_{Fr}(Q)\,\,\,.
\label{A20}
\end{equation}
\label{L1}
\end{lem}

Here, ${\det}_\zeta$ for $A$, $AQ$ are taken with respect to any
admissible spectral cuts close to $L_\theta$. The Fredholm determinant
${\det}_{Fr}$ is defined by
\begin{equation}
{\det}_{Fr}(\Id+S)=1+\Tr S+\Tr\wedge^2S+\ldots\,\,\,.
\label{A21}
\end{equation}
This series is absolutely convergent for any trace class operator $S$.
(Smoothing operators are of trace class.) Formula (\ref{A21})
is valid in a finite-dimensional case also.

The proof of (\ref{A20})
is based on applying a variation formula for an arbitrary smooth
$1$-parameter family $A_t$ of elliptic PDOs with
$\sigma\left(A_t\right)=\sigma(A)$, $A_0=A$, $A_1=AQ$.

The multiplicative anomaly $F(A,B)$ possesses a cocycle condition
$$
F(A,BC)F(B,C)=F(A,B)F(AB,C)
$$
(for any fixed spectral cuts for $A$, $B$, $C$, $AB$, $BC$, $ABC$).
We consider $F(A,B)$ as a ``partially defined and multi-valued
$2$-cocycle'' with the values in $\wC^{\times}$ on the group
$\SEll=G$ of elliptic symbols of index zero.
However, we can directly
construct the corresponding central $\wC^{\times}$-extension
$\sG$ of $G$. (Hence we do not work with a formalism of partially
defined cocycles.) The determinant Lie group $\sG$ is defined by formula
\begin{equation}
\sG=\sG(M,E):=\Ell^{\times}/H^{(1)},
\label{F1}
\end{equation}
where $H^{(1)}$ is the normal subgroup of the group $\Ell^{\times}$
of invertible elliptic PDOs of all complex orders,
$$
H^{(1)}=\left\{Q=\Id+S,\ S\text{ are smoothing, }{\det}_{Fr}Q=1\right\}.
$$

Note that the group $G$ of elliptic symbols takes the analogous form,
\begin{equation}
G=\Ell^{\times}/H,\quad H=\left\{Q=\Id+S,\ {\det}_{Fr}Q\ne 0\right\}.
\label{F2}
\end{equation}

There is a natural exact sequence
\begin{equation}
1\to\wC^{\times} @>>j> \sG @>>p> G\to 1\,\,\,.
\label{C6}
\end{equation}
Here, the identification $H/H^{(1)}\rs\wC^{\times}$ is given
by the Fredholm determinant (\ref{A21}). For any $A,B\in\Ell^{\times}$
we have
$$
d_1(A)d_1(B)=d_1(AB)\,\,\,,
$$
where $d_1\colon\Ell^{\times}\to\sG$ is the natural projection.
For a symbol $a\in G$, $\ord a=\alpha\in\wR^{\times}$, such that
the principal symbol $a_\alpha$ satisfies the Agmon-Nirenberg condition
with a sector $V$, we define a canonical element $d_0(a)\in\sG$,
$p\left(d_0(a)\right)=a$, by
\begin{equation}
d_0(a)=d_1(A)\cdot j\left(\left({\det}_\zeta(A)\right)^{-1}\right)\,\,.
\label{C5}
\end{equation}
Here, $A\in\Ell^{\times}$ is an arbitrary invertible elliptic PDO
with the symbol $a$, ${\det}_\zeta(A)$ is taken with respect to $V$.
Note that $j\left(\left({\det}_\zeta(A)\right)^{-1}\right)$ belongs
to the central subgroup $\wC^{\times}$ in $\sG$. The independence
$d_0(a)$ of $A$ (with $\sigma(A)=a$) follows immediately from (\ref{A20}).

Thus the multiplicative anomaly is enclosed in the central
$\wC^{\times}$-extension $\sG$ of $G$ with its partially defined
multi-valued section $d_0$ (over elliptic symbols of orders
from $\wR^{\times}\subset\wC$). Indeed,
$$
j(F(A,B))=d_0(\sigma(A))d_0(\sigma(B))d_0(\sigma(AB))^{-1}\,\,.
$$

Later on we use Lie algebras $\fell(M,E)$, $\frg=S_{\log}(M,E)$, $\sfrg$,
$\frh$, $\frh^{(1)}$ of all the Lie groups from above. The Lie algebra
$\fell(M,E)$ of the group $\Ell^{\times}$ consists of logarithms
of invertible elliptic PDOs and any element $l\in\fell(M,E)$ takes the form
$(q/2)\log(\Delta+\Id)+B$, where $q\in\wC$ and $B\in CL^0$. (Here,
$\Delta$ is the Laplacian for a Riemannian metric on $M$ and a unitary
connection on $E$.) The Lie algebra $\frg$ consists of the symbols
of elements from $\fell(M,E)$. These symbols are not classical.
In local coordinates on $M$ such a symbol takes the form
$q\log|\xi|\cdot\Id+b$, where $q\in\wC$ and $b$ is a zero order symbol.

Elements $l$ of $\frg$ are generators of one-parameter subgroups
$\exp(sl)$ of $\SEll=G$; $\left(\df_s\exp(sl)\right)\exp(-sl)=l$
in $\frg$.
Analogously, there are exponential maps from $\fell(M,E)$, $\frg$,
$\sfrg$, $\frh$, and $\frh^{(1)}$ to $\Ell^{\times}$, $G$, $\sG$,
$H$, and $H^{(1)}$.

\begin{rem}
The extension by $\log\xi$ (not by $\log|\xi|$) of the Lie algebra
of scalar Loran PDO-symbols of integer orders in the case of $M=S^1$
was considered in \cite{KrKh}. The authors of this paper also formally
constructed a central extension of the Lie algebra of such logarithmic
symbols with the help of the Adler-Manin-Lebedev residue. This cocycle is
analogous to one appearing on the right in formula (\ref{D23}),
with $x=\log\xi$. A multi-dimensional analog of this extension was
considered in \cite{R}. A connection of a formal Lie algebraic
construction of such a type with determinants of elliptic PDOs
investigated in \cite{KV} and here (Section~\ref{S6}), is a new fact.
\label{RH7}
\end{rem}

\section{New trace type functional}

Let $A\in CL^\alpha$ be a classical PDO of a noninteger order
$\alpha\in\wC\setminus\wZ$ acting on sections of a vector bundle $E$
on $M$, $\dim M=n$. We introduce a canonical density $t(A)$ on $M$
with the values in $\End(E)$ as follows. It is defined in any local
coordinate chart $U$ on $M$ together with a trivialization of $E$
over $U$. The density $t_U(A)$ is given by the restriction
to the diagonal $U\hookrightarrow U\times U$ of the difference
\begin{equation}
A(x,y)-\sum_{j=0}^NK_{-n-\alpha+j}(x,y-x)
\label{A16}
\end{equation}
of the Schwartz kernel $A(x,y)$ of $A$ (restricted to $U\times U$) and
the Fourier transforms of the first $N+1$, $N\gg 1$, homogeneous
terms $a_\alpha,a_{\alpha-1},\dots,a_{\alpha-N}$ of the symbol
$a=\sigma(A)$ with respect to given coordinates in $U$. Namely
$$
K_{-n-\alpha+j}(x,y-x):=(2\pi)^{-n}\int_{\wR_\xi^n}\exp(i(x-y,\xi))a_{\alpha-j}
(x,\xi)d\xi\,\,\,.
$$
This distribution is positive homogeneous in $y-x\in\wR^n$ of order
$-n-\alpha+j$ for $\alpha\notin\wZ$.

Note that any positive homogeneous
distribution from $\cD'\left(\wR^n\setminus 0\right)$ of order
$\beta\notin\{m\in\wZ,m\le-n\}$ has a unique prolongation to a positive
homogeneous distribution from $\cD'\left(\wR^n\right)$ (see \cite{Ho},
Theorem~3.2.3). Hence, if we restrict $K_{-n-\alpha+j}(x,y-x)$ to $y\ne x$,
we'll not lose any information.

\begin{lem}
The difference (\ref{A16}) is a continuous on $U\times U$ function
for $N$ large enough. Hence its restriction $t_U(A)$ to the diagonal $U$
makes sense.
\label{LB1}
\end{lem}

\begin{lem}
The density $t_U(A)$ with the values in $\End E$ is independent of large $N$,
of local coordinates on $M$, and of a local trivialization of $E$.
\label{LB2}
\end{lem}

The statement of Lemma~\ref{LB1} follows directly from the structure
of singularities of PDO-kernels. The independence $t_U(A)$ of the change
$N$ by $N+1$ (if $N$ is large enough) follows from the positive homogeneity
of $K_{-n-\alpha+N+1}(x,y-x)$ in $y-x$ and from the fact
$\Re(-n-\alpha+N+1)>0$. The invariance of $t_U(A)$ under changes
of local coordinates and of trivializations follow from the Taylor's
formula, from the non-integrality of $\alpha$, and from ordinary
properties of derivatives of homogeneous functions.

\begin{thm}
The linear functional
$$
\TR(A)=\int_M\tr t(A)
$$
on classical PDOs of orders from $\alpha_0+\wZ$,
$\alpha_0\in\wC\setminus\wZ$, in the case of a closed $M$ has the following
properties.

1. It coincides with the usual trace $\Tr A$ in $L_2(M,E)$
for $\Re\ord A<-n$.

2. It is a trace type functional, i.e., $\TR([B,C])=0$
for $\ord B+\ord C\in\alpha_0+\wZ$.

3. For any holomorphic family $A(z)$ of classical PDOs on $M$,
$z\in U\subset\wC$, $\ord A(z)=z$, the function $\TR(A(z))$ is
meromorphic with no more than simple poles at $z=m\in U\cap\wZ$ and
with residues
\begin{equation}
\Res_{z=m}\TR(A(z))=-\res\sigma(A(m))\,\,.
\label{A25}
\end{equation}
(Here, $\res$ is the noncommutative residue of the symbol of $A(m)$,
$m\in\wZ$, \cite{Wo1}, \cite{Kas}, \cite{Wo2}.)
\label{TB3}
\end{thm}

The part 2. follows from the parts 1. and 3. applied to arbitrary
holomorphic families $B(z)$, $C(z)$, $z\in U$, such that $B(0)=B$,
$C(0)=C$, and $B(z)$, $C(z)$ are of trace class in some subdomain
$U_1\subset U$ (i.e., $\Re\ord B(z),\Re\ord C(z)<-n$ for $z\in U_1$).

In the part 3., (\ref{A25}), we use that the singularities of densities
$t(A(z))$ are the same as of the restriction to the diagonal
of the integral
\begin{equation}
\sum_{j=0}^N\int(\rho(|\xi|)-1)|\xi|^{z-j}a_{z-j}(z,x,\xi/|\xi|)
\exp(i(x-y,\xi))d\xi\,\,\,.
\label{H10}
\end{equation}
Here, $\rho(|\xi|)$ is a smooth cutting function, $\rho(|\xi|)=1$
for $|\xi|\gg 1$, $\rho(|\xi|)=0$ for $|\xi|\ll 1$.
The integral (\ref{H10}) for $x=y$ has an explicit analytic continuation
produced with the help of the the equality
$\int_0^1x^\lambda dx=1/(\lambda+1)$, $\Re\lambda>-1$.

\begin{rem}
Theorem~\ref{TB3} implies that $\res([b,c])=0$ for $\ord b+\ord c\in\wZ$
(i.e., $\res$ is a trace type functional). This assertion is well
known, \cite{Kas}, \cite{Wo2}, but its usual proof is not so elementary
because it uses the spectral interpretation of the noncommutative
residue.
\label{RA30}
\end{rem}

\section{Applications to zeta-functions}

The trace type functional $\TR$ introduced in the previous section
gives us a tool to define zeta-functions for one-parameter subgroups
of $\Ell^{\times}$ denerated by elements $x\in\fell(M,E)$ with
$\ord(\exp x)\ne 0$. From now on we denote $\ord(\exp x)$ by $\ord x$
for any $x\in\fell(M,E)$.
 We define
$$
\zeta_x^{\TR}(s):=\TR\exp(-sx)
$$
for $s\ord x\notin\wZ$. By Theorem~\ref{TB3} we conclude the following.
\begin{pro}
1. The zeta-function $\zeta_x^{\TR}(s)$ is a meromorphic function
on $\wC\ni s$ with at most simple poles at $s_k=k/\ord x$, $k\in\wZ$,
$k\le n:=\dim M$. This function is regular at $s=0$ by (\ref{A25}).

2. The residue of $\zeta_x^{\TR}(s)$ at $s_k$ is
\begin{equation*}
\Re_{s=s_k}\zeta_x^{\TR}(s)=\res\sigma\left(\exp\left(-s_kx\right)\right)/
\ord x\,\,\,.
\end{equation*}

3. Let $\ord x\in\wR^{\times}$ and let $\exp x$ possess a spectral cut
$L_\theta$ such that the $\log\exp x$ defined with respect to $L_\theta$
is equal to $x$. Then we have
$$
\zeta_x^{\TR}(s)=\zeta_{\exp x,\theta}(s)\,\,\,,
$$
i.e., in this case, $\zeta_x^{\TR}(s)$ coincides with the classical
zeta-function.
\label{PB5}
\end{pro}

Note that the functional $\TR$ gives us a tool to define zeta-functions
without an analytic continuation. For example, our definition has
an immediate consequence, which is out of reach of previous methods.

\begin{cor}
Let $A^s$ and $B^s$ be two holomorphic families of complex powers
such that $A^{s_0}=B^{t_0}$ and let $\ord A\cdot s_0\notin\wZ$.
Then
$$
\zeta_A\left(s_0\right)=\zeta_B\left(t_0\right)\,\,\,,
$$
where zeta-functions are defined by the meromorphic continuation
from the domains of convergence.
\label{CB9}
\end{cor}

Theorem~\ref{TB3} provides us with a general information on the structure
of derivatives of zeta-functions at zero.

\begin{thm}
There are homogeneous polynomials $T_{k+1}(x)$ of order $k+1\ge 1$
in $x\in\fell(M,E)$ such that
$$
\ord x\cdot\df_s^k\zeta_x^{\TR}(s)|_{s=0}=T_{k+1}(x)\,\,\,.
$$
These polynomials are invariant with respect to the adjoint action
of $\Ell^{\times}$ on $\fell(M,\!E)$. The restriction of $T_{k+1}(x)$
to the Lie ideal (of codimension one) $CL^0=\{x\colon\ord x=0\}$ is
\begin{equation}
T_{k+1}(x)\big|_{\ord x=0}={(-1)^{k+1}\over(k+1)}\res\left(\sigma(x)^{k+1}
\right)\big|_{\ord x=0}\,\,\,.
\label{B15}
\end{equation}
\label{TB10}
\end{thm}

The statement 3. of Theorem~\ref{TB3} applied to a holomorphic family
$\exp(sx+b)$, $\ord x=1$, $b\in CL^0$, near $s=0$ implies that
the function $\ord x\cdot\TR\exp(x)$ is holomorphic at $\ord x=0$
(on $\fell\ni x$). The polynomials $T_k(x)$ are (up to standard factors)
the Taylor coefficients of this function at $x=0$.

\begin{cor}
The function $\log\det(\exp x)$ is the ratio $-T_2(x)/\ord x$
of a quadratic function and a linear function. But it is not
a linear function (by (\ref{B15})). There is no linear function
``$\,\Tr$'' on $\fell$ such that ``$\,\Tr$''$\,\log A=\log\det(A)$.
\label{CB16}
\end{cor}

A statement analogous to Proposition~\ref{PB5} holds also for a generalized
zeta-function
$\TR\left(B_1A_1^{s_1}\ldots B_kA_k^{s_k}\right)=:\zeta_{(A_j),(B_j)}\left(
s_1,\dots,s_k\right)$. Here, $A_j^{s_j}$ are holomorphic families of powers
of elliptic PDOs $A_j$ (not all of $\alpha_j:=\ord A_j$ are equal to zero),
$s_j\in\wC$, and $B_j$ are classical PDOs of orders $\beta_j$.

\begin{pro}
The zeta-function $\zeta_{(A_j),(B_j)}\left(s_1,\dots,s_k\right)$ is
meromorphic in $\bs:=\left(s_1,\dots,s_k\right)\in\wC^k$ with at most
simple poles on the hyperplanes
$z(\bs)\!:=\!\sum_j\left(\beta_j\!+\!s_j\alpha_j\right)\!=m\in\wZ$, $m\ge-n$.
Its residue is equal
to $-\res\sigma\left(B_1A_1^{s_1}\ldots B_kA_k^{s_k}\right)|_{z(\bs)=m}$
and thus it is computable in terms of symbols of $B_j$ and of $\log A_j$.
\label{PB22}
\end{pro}

\section{Canonical determinant}

In this section we return to the determinant Lie group $G$. Above
we have constructed, (\ref{C5}), the multi-valued section $d_0$
of the $\wC^{\times}$-bundle $\sG\to G$. Here we extend $d_0$ to its
maximal natural domain of definition and introduce (with the help
of extended $d_0$) the canonical determinant of elliptic PDOs.

Let $a\in G:=\SEll$ be an elliptic symbol of a nonzero order and let
$a=\exp x$ for some $x\in\frg=S_{\log}$.
Pick any $b\in\fell(M,E)$ such that its symbol $\sigma(b)$ is $x$.
Then we define $d_0(a,x)$ as
\begin{equation}
d_0(a,x):=d_1(\exp b)j\left(\exp\left(\df_s\zeta_b^{\TR}(s)|_{s=0}\right)
\right)\in\sG\,\,\,.
\label{C0}
\end{equation}

\begin{lem}
The element $d_0(a,x)$ is independent of $b\in\fell(M,E)$ with
$\sigma(b)=x$.
\label{LC1}
\end{lem}

This lemma together with its proof is analogous to Lemma~\ref{L1}.

\begin{rem}
The element $d_0(a,x)$, (\ref{C0}), depends on $x\in S_{\log}$ only,
because $a=\exp x$. Also $d_0(a,x)$ is analytic in $x$, $\ord x\ne 0$.
Elements $d_0(a,x)$ for $\ord a\in\wC^{\times}$ form the image
under the exponential map of a $\wC^{\times}$-cone in the Lie algebra
$\sfrg$. In the next section we prove that this cone is a quadratic one
and give an explicit description of it in terms of symbols.
\label{RC9}
\end{rem}

\begin{lem}
For $a$, $\ord a\in\wR^{\times}$, possessing a spectral cut $L_\theta$
and such that the logarithm of $a$ (with respect to $L_\theta$) is equal
to $x$, the definitions (\ref{C0}) and (\ref{C5}) coincide.
\label{LC2}
\end{lem}

\noindent{\bf Definition}. Let $A\in\Ell^{\times}$ be of any
nonzero complex order.
Let its symbol $a:=\sigma(A)$ have a logarithm $x\in S_{\log}$.
Then the {\em canonical determinant} of $A$ is defined as
\begin{equation}
\det(A,x):=j^{-1}\left(d_1(A)d_0(a,x)^{-1}\right)\in\wC^{\times}\,\,.
\label{C7}
\end{equation}
(Here, $j\colon\wC^{\times}\hookrightarrow\sfrg$ is the natual inclusion
of the central subgroup from (\ref{C6}).)

\begin{rem}
This definition does not use any family $A^s$ of complex powers of $A$.
It uses families of powers $\exp(sb)=(\exp(b))^s$
of $b$ with $\sigma(b)=\log\sigma(A)$ constructed elementary by any such $b$.
The necessity of such a construction with powers of other operators
follows from the fact that the existence of a logarithm of a generic
invertible elliptic PDO cannot be described in terms of his symbol.
Also the image of the exponential map
$\exp\colon S_{\log}\to\SEll=G$ has much more simple structure
than the image of $\exp\colon\fell\to\Ell^{\times}$. (See more detailed
discussion of this problem in \cite {KV}, Remarks~6.3, 6.4, 6.8, 6.9.)
In (\ref{C7}) we use only the existence of $\log\sigma(A)$.
\label{RC8}
\end{rem}

\subsection{Microlocal Agmon-Nirenberg condition}
\label{S5.1}

Here we introduce a sufficient condition of the existence
of $\log\sigma(A)$ generalizing the Agmon-Nirenberg condition.
Let $a_\alpha(x,\xi)$ be the principal elliptic symbol of $A$,
$\alpha=\ord A\in\wC^{\times}$.
Let $\bold{\theta}:=\theta(x,\xi)\colon T^*M\setminus M\to\wR$
be a continuous map such that
$L_{\theta(x,\xi)}\cap\Spec a_\alpha(x,\xi)=\emptyset$ for all
$(x,\xi)\in T^*M\setminus M\,\,$.

\begin{lem}
Under this condition, $\log\sigma(A)$ exists. It is explicitly defined
by the formula analogous to (\ref{A15}) on the level of complete
symbols. Namely $\log_{(\theta)}\sigma(A)(x,\xi)$ is the derivative
at $s=0$ of the family of symbols $\sigma(A)_\theta^s$.
Here, $\sigma(A)_\theta^{-s}$ is defined for $\Re s>0$ by the integral
\begin{equation}
{i\over 2\pi}\int_{\Gamma_{\theta(x,\xi)}}(\sigma(A)-\lambda)^{-1}(x,\xi)
\lambda_{(\theta(x,\xi))}^{-s}d\lambda\,\,\,,
\label{C15}
\end{equation}
and $(\sigma(A)-\lambda)^{-1}$ is an inverse element in the algebra
of symbols (with a parameter $\lambda$ of homogeneity degree $\alpha$).
For $-k<\Re s\le 0$, $k\in\wZ_+$,
$\sigma(A)_\theta^{-s}:=\sigma(A)^k\cdot\sigma(A)_\theta^{-s-k}$.
\label{LC14}
\end{lem}

\begin{rem}
The definition (\ref{C15}) of $\sigma(A)_\theta^s$ is invariant
under homotopies of a field $\theta(x,\xi)$
of admissible for $a_{\alpha}$ cuts. By the homotopy equivalence
$S^*M\sim T^*M\setminus M$ and positive homogeneity of $a_\alpha(x,\xi)$
it is enough to define $\bold{\theta}$ only over any global smooth section
of the $\wR^{\times}_{+}$-bundle
$T^*M\setminus M\to S^*M$. The existence of a field of admissible
for $a_\alpha(x,\xi)$ cuts is in a sense nonsensitive to an order
$\alpha\in\wC^{\times}$. It is applicable to elliptic symbols
of complex orders.
\label{RF1}
\end{rem}

\begin{rem}
The microlocal Agmon-Nirenberg condition of Lemma~\ref{LC14} is a rather
weak restriction on $\sigma(A)$.
Nevertheless there are simple topological obstructions to the existence
of $\log\sigma(A)$. For instance, for any $(M,E)$ with $\dim M\ge 2$,
$\rk E\ge 2$, there are nonempty open subsets in $\Ell^{\times}(M,E)$
admitting no continuous logarithms of principal elliptic symbols.
For example, let the principal symbol $a_\alpha(x,\xi)$ have
at $\left(x_0,\xi_0\right)$ a Jordan block
$\left
(\begin{matrix}
\lambda & 1 \\
0       & \lambda
\end{matrix}
\right)$. Let the corresponding to $\lambda$ eigenvalues over a closed
curve $S^1\to S^*M$ be $\lambda_i(\phi)$,
$\lambda_i\left(\phi_0\right)=\lambda$, $i=1,2$, and let the winding
numbers of $\lambda_i(\phi)$ be $+m$, $-m$, where $m\in\wZ\setminus 0$.
Then there is no continuous $\log a_\alpha(x,\xi)$. This condition
is an open condition on a principal symbol.
\label{RD10}
\end{rem}

\section{Determinant Lie algebra and quadratic cone}
\label{S6}

We know by Theorem~\ref{TB10} that the first derivative of the zeta-function
at zero is given by
$$
\df_s\zeta_x^{\TR}(s)|_{s=0}=T_2(x)/\ord x\,\,\,.
$$
Here, $T_2$ is a quadratic form on $\fell(M,E)\ni x$. The associated
symmetric bilinear form $B$ on $\fell$,
$$
B(x,y):=T_2(x+y)-T_2(x)-T_2(y)\,\,\,,
$$
has the following properties.

\begin{lem}
1. $B(x,y)$ is invariant under the adjoint action,
$$
B([x,z],y)+B(x,[y,z])=0
$$
for $x,y,z\in\fell$.

2. For $y\in\frh\subset\fell$ we have
$$
B(x,y)=-\Tr y\cdot\ord x\,\,\,.
$$
(operator $y\in\frh$ is smoothing and hence is of trace class.)

3. For $x,y\in CL^0\subset\fell$ we have
$$
B(x,y)=(\sigma(x),\sigma(y))_{\res}:=\res(\sigma(x)\sigma(y))\,\,\,.
$$
\label{LD1}
\end{lem}

The properties 1. and 3. of $B(x,y)$ follow immediately
from Theorem~\ref{TB10}, (\ref{B15}). The property 2. is a consequence
of the equality
\begin{equation}
\df_t\left(\df_s\zeta_{x_t}^{\TR}(s)|_{s=0}\right)=-\Tr\left(\int_0^1ds\cdot
\Ad_{\exp(sx_t)}\df_tx_t\right)=-\Tr\left(\df_tx_t\right)\,\,.
\label{D2}
\end{equation}
Here, $x_t:=x+ty$, $\ord x\ne 0$. (In (\ref{D2}) we use that
$\df_tx_t=y$ is a smoothing operator.)

The property 2. implies that $\frh^{(1)}\subset\Ker B$. Hence $B$ induces
an invariant symmetric bilinear form on $\sfrg:=\fell/\frh^{(1)}$. We denote
this form by the same letter $B$.

\begin{pro}
1. For any $x\in\frg=S_{\log}$, $\ord x\ne 0$, there exists a unique
$\sx\in\sfrg$ such that $p\sx=x$ ($p\colon\sfrg\to\frg$ is the natural
projection) and such that $B\left(\sx,\sx\right)=0$ (i.e., $\sx$ is
an isotropic vector).

2. The element $d_0(\exp x,x)$ defined by (\ref{C0}) (for $\ord x\ne 0$)
is given by
\begin{equation*}
d_0(\exp x,x)=\exp\left(\sx\right)\,\,.
\end{equation*}
\label{PD3}
\end{pro}

The part 1. follows from the condition $\ord x\ne 0$ because
$B(x,j(1))=-\ord x$ (see Lemma~\ref{LD1}, {\em 2.}).
Here, $1\in\frh/\frh^{(1)}\simeq\wC$ is represented by any smoothing
operator with the trace equal
to $1$ and $j\colon\frh/\frh^{(1)}\hookrightarrow\sfrg$ is the natural
inclusion.

The part 2. follows from the equality
\begin{equation*}
\df_s\zeta_b^{\TR}(s)|_{s=0}={B(\sx,\sx)\over 2\ord x}=0
\end{equation*}
for any $b\in\fell$ such that $\sx=b\left(\mod\frh^{(1)}\right)$
in $\sfrg=\fell/\frh^{(1)}$.

Now we describe a general algebraic construction which in our case
gives the description of the determinant Lie algebra $\sfrg$ (and
of the bilinear form $B$ on $\sfrg$) in terms of symbols.

Let us consider a central extension
\begin{equation}
0\to\wC @>>j>\sfrg @>>p>\frg\to 0
\label{D20}
\end{equation}
of an abstract Lie algebra $\frg$ with an invariant symmetric bilinear form
$B$ on $\sfrg$ such that $B(j(1),j(1))=0$ and $B(j(1),x)\not\equiv 0$
(i.e., $\Im j\not\subset\Ker B$). We associate with (\ref{D20})
an exact sequence of Lie algebras
\begin{equation}
0\to\frg_0 @>>r>\frg @>>q>\wC\to 0\,\,\,.
\label{D21}
\end{equation}

Here, $qx:=-B\left(j(1),x_1\right)$, $x\in\frg$, for any $x_1\in p^{-1}(x)$,
and $\frg_0:=\Ker q$ is a codimension one Lie ideal in $\frg$.
The form $B$ on $\sfrg$ induces a symmetric bilinear form $B_0$
on $\frg_0$. Namely, $B_0(x,y):=B\left(x_1,y_1\right)$ for any
$x_1\in p^{-1}(r(x))$, $y_1\in p^{-1}(r(y))$. The form $B_0$ is
invariant under the adjoint action of $\frg$ on $\frg_0$.

In our concrete situation $\frg_0$ is $CL^0$, $\frg$ is $S_{\log}$,
$q(x)=\ord x$, $r$ is the natural inclusion, and $B_0(x,y)=(x,y)_{\res}$.

\begin{thm}
The exact sequence (\ref{D20}) and the symmetric bilinear form $B$ on $\sfrg$
can be canonically reconstructed from (\ref{D21}) and from the form $B_0$
on $\frg_0$.
\label{TD22}
\end{thm}

\noindent{\bf Proof.} 1. Suppose we have both sequences, (\ref{D20}) and
(\ref{D21}), related one with another as described above. Then for any
$x\in\frg$ with $q(x)=1$ we have a unique $\sx\in\sfrg$, $B(\sx,\sx)=0$,
$p(\sx)=x$. The hyperplane $\{y\in\sfrg:\,B(\sx,y)=0\}\subset\sfrg$
defines the splitting of (\ref{D20}) (as of the exact sequence of vector
spaces) $\Pi_{\sx}\colon\frg\hookrightarrow\sfrg$. One can check that
the Lie bracket and the form $B$
on $\sfrg\simeq\Pi_{\sx}(\frg)\oplus j(\wC)$ are as follows.
\begin{multline}
\left[\Pi_{\sx}\left(r\left(a_1\right)+t_1x\right)+j\left(c_1\right),\Pi_{\sx}
\left(r\left(a_2\right)+t_2x\right)+j\left(c_2\right)\right]= \\
=\Pi_{\sx}\left(\left[r\left(a_1\right)+t_1x,r\left(a_2\right)+t_2x\right]
\right)-j\left(B_0\left(r^{-1}\left[x,r\left(a_1\right)\right],a_2\right)
\right),
\label{D23}
\end{multline}
\begin{multline}
B\left(\Pi_{\sx}\left(r\left(a_1\right)+t_1x\right)+j\left(c_1\right),
\Pi_{\sx}\left(r\left(a_2\right)+t_2x\right)+j\left(c_2\right)\right)= \\
=B_0\left(a_1,a_2\right)-c_1t_2-c_2t_1\,\,\,.
\label{D24}
\end{multline}
Here we use the parametrization $\Pi_{\sx}(r(a)+tx)+j(c)$, $a\in\frg_0$,
$t,c\in\wC$, of $\sfrg$ (and the parametrization $r(a)+tx$ of $\frg$).

Let we have two elements $x$, $x'$ in $\frg$ such that $q(x)=q(x')=1$.
Then we have
\begin{equation}
\left(\Pi_{\sx}-\Pi_{\sx'}\right)(y)=B_0\left(r^{-1}(y-q(y)(x+x')/2),r^{-1}
(x-x')\right)\cdot j(1)\,\,\,.
\label{D25}
\end{equation}

2. Formulas (\ref{D23}), (\ref{D24}) can be interpreted as a construction
of the extension (\ref{D20}) and of the symmetric bilinear form $B$ on $\sfrg$
in terms of (\ref{D21}) and $B_0$. This construction
of $(\sfrg,B)\simeq\left(\sfrg_x,B_x\right)$ depends on a choice
of $x\in q^{-1}(1)$. Formula (\ref{D25}) provides us with an associative
system of identifications of these Lie algebras $\sfrg_x$ (together
with the bilinear forms $B_x$ on them) for different $x$.\ \ \ $\Box$

\noindent{\bf Conclusions.} 1. We obtain a description
(\ref{D23})--(\ref{D25}) of the determinant Lie algebra $\sfrg$ in terms
of $\frg=S_{\log}$, i.e., in terms of symbols (without using Fredholm
determinants and so on). Namely, $\sfrg$ is generated by symbols
$\Pi_x y$ for $x\in q^{-1}(1)\subset\frg$, $y\in\frg$, and by $j(c)$,
$c\in\wC$. Symbols $\Pi_x y$ are linear in $y$, and $j(c)=cj(1)$.
The relations between these symbols are given by (\ref{D23}),
(\ref{D25}), where $\Pi_{\sx}$ is replaced by $\Pi_x$ and where
$y$ is represented by $r(a)+tx$. Formula (\ref{D24}) defines
an invariant bilinear form on $\sfrg$.

2. Let $A$ be an elliptic PDO, $\ord A\ne 0$, with a given
$\log\sigma(A)=:x\in S_{\log}=\frg$. Then we have an explicit formula
for $\log d_0(\sigma(A),x)$ in $\sfrg$. Namely
\begin{equation*}
\log d_0(\sigma(A),x)=\Pi_{(\widetilde{x/\ord x})}(x)\,\,\,.
\end{equation*}

\begin{rem}
The central extension $\sfrg$ is not a trivial extension of the topological
Lie algebra $\frg$. This fact can be proved with the help
of the Atiyah-Singer Index theorem for families of elliptic PDOs.
See Lemma~6.8 in \cite{KV}.
\label{RH12}
\end{rem}

\subsection{Singularities of determinants}

Let $A(z)$ be a holomorphic family of invertible elliptic PDOs,
$z\in U\subset\wC$.
Let for a one-connected subdomain $U_1\subset\overline{U_1}\subset U$
a family $\log\sigma(A(z))$, $z\in U_1$, be defined (e.g., using microlocal
fields of spectral cuts as in Section~\ref{S5.1} depending on $z$). We are
interested in analytic behaviour of $\det(A(z))$ near $\df U_1$.
Using previous constructions we can find a function $f(z)$
 from $U_1\ni z$ to $\wC^{\times}$, such that \\
1. $f(z)$ is defined by $\log\sigma(A(z))$, \\
2. $\det(A(z),\log\sigma(A(z)))/f(z)$ is holomorphic in a neighborhood
of $\overline{U_1}$ in $U$. \\

\noindent{\bf Construction of $f(z)$.} Fix a logarithmic symbol
$x\in S_{\log}$
of order $1$. It determines a splitting of the determinant Lie algebra
$\sfrg$ via the map $\Pi_{\sx}$. (See Proof of Theorem~\ref{TD22}.)
So it defines a right invariant holomorphic connection
on the $\wC^{\times}$-bundle $\sG @>>p> G$. Let $d_2(z)$ be a flat section
of the pullback under the map $z\to\sigma(A(z))$ of this determinant
bundle $p$ on a neighborhood of $\overline{U_1}$.
Set $f(z):=d_2(z)/d_0(\sigma(A(z)),\log\sigma(A(z)))$, $z\in U_1$.

\begin{pro}
$\det(\!A(\!z\!),\log\sigma(\!A(\!z\!)))/\!f(\!z\!)$ has a holomorphic
extension to a neighborhood of $\overline{U_1}$.
\label{PB57}
\end{pro}

This assertion is clear because
$$
\det(A(z),\log\sigma(A(z)))/f(z)=d_1(A(z))/d_2(z)\,\,,
$$
and the both factors on the right are holomorphic
in a neighborhood of $\overline{U_1}$.

\section{Odd-dimensional case}

The algebra of classical elliptic PDOs contain an invariantly defined
subalgebra of odd class operators.

\noindent{\bf Definition.} Let $\left\{U_i\right\}$ be a cover of $M$
by coordinate charts and let $E|_{U_i}$ be trivialized. Then
$A\in CL^d$, $d\in\wZ$, is an {\em odd class} PDO if its symbol
$\sigma(A)$ obeys on any $U_i$ the following condition
\begin{equation*}
\sigma_k(A)(x,\xi)=(-1)^k\sigma_k(A)(x,-\xi)\,\,\,.
\end{equation*}

Here, $\sigma_k(A)$, $k\le d$, are positive homogeneous in $\xi$
components of $\sigma(A)$ in charts $U_i$. This condition is independent
of a choice of local coordinates (near $x\in M$) and of a trivialization
of $E$. It follows from the transformation formula for PDO-symbols
under changing of space coordinates.

We denote by $CL_{(-1)}^{\wZ}$ the linear space of odd class PDOs, and
by $CS_{(-1)}^{\wZ}$ the space of their symbols.
By $\Ell_{(-1)}^{\times}$ we denote the group of odd class invertible
elliptic PDOs and by $\SEll_{(-1)}$ the group of their symbols.
 $\Ell_{(-1)}^{\times}$ and $\SEll_{(-1)}$ are groups by the following Lemma.

\begin{lem}
1. Differential operators (DOs) are contained in $CL_{(-1)}^{\wZ}$.

2. Smoothing operators are contained in $CL_{(-1)}^{\wZ}$.

3. $CL_{(-1)}^{\wZ}$ is a subalgebra of $CL^{\wZ}$.

4. If $A\in \Ell_{(-1)}^{\wZ}$ is an invertible elliptic PDO, then
$A^{-1}\in \Ell_{(-1)}^{\wZ}$.
\label{LE2}
\end{lem}

\begin{pro}
Let $a\in\SEll_{(-1)}^{2k}$, $k\in\wZ$, be an elliptic symbol admitting
a {\em microlocal field $\theta$ of spectral cuts}
(Section~\ref{S5.1}), which is {\em projective},
i.e., $\theta(x,\xi)=\theta(x,-\xi)$ for $\xi\ne 0$. Then we have

1. $a_{(\theta)}^{1/k}\in\SEll_{(-1)}^2$ for $k\ne 0$.

2. $a_{(\theta)}^s$ belongs to $\SEll_{(-1)}^0$
for $k=0$.

3. $\log_{(\theta)}a$ belongs to $CS_{(-1)}^0$ for $k=0$.
\label{PE5}
\end{pro}

\begin{lem}
We have $\,\,\,\exp a\in\SEll_{(-1)}^0$ for $a\in CS_{(-1)}^0$.
\label{LF10}
\end{lem}

\begin{pro}
Let elliptic symbols $a_1\in\SEll_{(-1)}^{2k_1}$ and
$a_2\in\SEll_{(-1)}^{2k_2}$,
$k_j\in\wZ\setminus 0$, admit projective fields of spectral cuts
$\theta_1$ and $\theta_2$. Then
\begin{equation}
{\log_{(\theta_1)}a_1\over k_1}-{\log_{(\theta_2)}a_2\over k_2}\in CS_{(-1)}
^0\,\,\,.
\label{E7}
\end{equation}
\label{PE6}
\end{pro}

For $a$ and $\theta$ as in Proposition~\ref{PE5} we denote
by $a_{(\theta),2ks-j}^s(x,\xi)$, $j\in\zuo$, the homogeneous components
of $a_{(\theta)}^s$. Then assertions of Propositions~\ref{PE5} and
\ref{PE6} follow from the equalities
\begin{align}
a_{(\theta),2ks-j}^s(x,\xi) & =(-1)^ja_{(\theta),2ks-j}^s(x,-\xi)\,\,\,,
\label{E8} \\
a_{(\theta)}^0              & =\Id\,\,\,. \notag
\end{align}
The equality (\ref{E8}) is a direct consequence of the integral
representation (\ref{C15}) for $a_{(\theta)}^s$ and of explicit
formulas for the symbol $(a-\lambda)^{-1}$ (with $\deg\lambda=\ord a$).

{}From now on we suppose that $M$ is odd-dimensional.

\begin{lem}
For $A\in CL_{(-1)}^{\wZ}$ we have
\begin{equation}
\res\sigma(A)=0\,\,\,.
\label{E4}
\end{equation}
\label{LE3}
\end{lem}

This formula follows immediately from the definition of $\res$
because $\sigma_{-n}(A)(x,\xi)$ is odd in $\xi$, $n:=\dim M$.

Now we have tools for investigation of the multiplicative anomaly
in the odd class. Let $A$ and $B$ be odd class invertible elliptic
PDOs of nonzero even orders, $\ord A+\ord B\ne 0$, such that the symbols
of $A$, $B$, and of $AB$ admit projective fields of spectral cuts
$\theta_1$, $\theta_2$, $\theta_3$. The multiplicative anomaly is
defined (in this case) as
$$
F(A,B):={\det\left(AB,\log_{(\theta_3)}\sigma(AB)\right)\over\det\left(A,
\log_{(\theta_1)}\sigma(A)\right)\det\left(B,\log_{(\theta_2)}\sigma(B)
\right)}\,\,\,.
$$

\begin{thm}
1. $F(A,B)$ is locally constant in $A$, $B$ (for given admissible
$\theta_j$).

2. For the principal symbols of $A$ and $B$ sufficiently close to positive
definite self-adjoint ones and for fields $\theta_j$ close to $\pi$
we have
\begin{equation}
F(A,B)=1\,\,\,.
\label{E10}
\end{equation}
\label{TE9}
\end{thm}

The multiplicative property (\ref{E10}) for zeta-regularized determinants
of positive self-adjoint deifferential operators on closed odd-dimensional
manifolds is a new one.

Our proof of Theorem~\ref{TE9} is based on a general variation formula
analogous to (\ref{A46}) (valid without assumptions that
$A,B$ are of odd class and that $\dim M$ is odd)
\begin{equation*}
{\df\over\df t}\log F\left(A_t,B\right)=
-\left(\sigma\left({\df\over\df t}A_t\cdot A_t^{-1}\right),{\log_{(\theta_3)}
\sigma(AB)\over\ord A+\ord B}-{\log_{(\theta_1)}\sigma(A)\over\ord A}\right)
_{\res}.
\end{equation*}

According to Proposition~\ref{PE6} and to Lemmas~\ref{LE2}, \ref{LE3}
the right hand side is equal to zero.

The multiplicative property (\ref{E10}) provides us with a possibility
to define $\det(A)$ for any invertible elliptic $A$ from $\Ell_{(-1)}^0$
close to positive definite self-adjoint ones. Namely, define $\det A$
as
\begin{equation}
\det(A):={\det}_\zeta(AB)/{\det}_\zeta(B)
\label{E15}
\end{equation}
for an arbitrary positive self-adjoint invertible $B\in\Ell_{(-1)}^{2k}$,
$k>0$. Here, spectral cuts are close to $\wR_-$. Independence
of the expression on the right in (\ref{E15}) of $B$ follows
from the equality (\ref{E10}).

\begin{lem}
For an elliptic DO $A$ of zero order (i.e., for $A\in\Aut E$) sufficiently
close to positive definite ones, its determinant (\ref{E15}) is equal
to $1$.
\label{LF14}
\end{lem}

This statement can be proved by the remark that the map
$$
q\colon\End E\ni f\to\df_s\log\det(\exp(sf))|_{s=0}\in\wC
$$
is a homomorphism of Lie algebras. Here, $\det$ is defined by (\ref{E15}).
The map $q$ is invariant under the adjoint action of the group
$\Diff(M,E)$ of diffeomorphisms of the total space of $E$
which are linear maps between fibers. It is clear that the only
$\Diff(M,E)$-invariant continuous linear functional on $\End(E)$
is $q\equiv 0$.

\subsection{Determinant Lie group for odd class operators}

We define the Lie groups
$$
G_{(-1)}:=\Ell_{(-1)}^{\times}/H=\SEll_{(-1)},\quad \sG_{(-1)}:=\Ell_{(-1)}
^{\times}/H^{(1)}\,\,\,,
$$
analogous to (\ref{F2}), (\ref{F1}). We call $\sG_{(-1)}$
the determinant Lie group for the odd class. (To remind, $M$ is a closed
odd-dimensional manifold.)

By Proposition~\ref{PE5} the Lie algebra $\frg_{(-1)}$ of $G_{(-1)}$ is
$CS_{(-1)}^0$. The group $\sG_{(-1)}$ is a central $\wC^{\times}$-extension
of $G_{(-1)}$ by analogy with (\ref{C6}).
 Here again, the identification $\wC^{\times}=H/H^{(1)}$ is defined
by $\det_{Fr}$.

\begin{pro}
The Lie algebra $\sfrg_{(-1)}$ of $\sG_{(-1)}$ is canonically splitted
into the direct sum of the Lie algebras
\begin{equation}
\sfrg_{(-1)}=\wC\oplus\frg_{(-1)}\,\,\,.
\label{F5}
\end{equation}
This splitting is invariant with respect to the adjoint action
of $\sG_{(-1)}$ on $\sfrg_{(-1)}$.
\label{PF4}
\end{pro}

For the splitting (\ref{F5}) we can use any elliptic symbol
$a\in\SEll_{(-1)}^{2k}$, $k\in\wZ\setminus 0$, admitting a projective
microlocal field of spectral cuts $\theta$.
Set $x:=(1/2k)\log_{(\theta)}a\in\frg$. We embed $\frg_{(-1)}$ into
$\sfrg_{(-1)}$ by the map $\Pi_{\sx}$. (See Proof of Theorem~\ref{TD22}.)
We claim: \\
1. $\Pi_{\sx}$ is a Lie algebra homomorphism. \\
2. $\Pi_{\sx}$ is independent of $a$, $\theta$.

The first assertion follows from Lemma~\ref{LE2}, {\em 3}, Lemma~\ref{LE3},
and from formula (\ref{D23}). The second assertion is a consequence
of (\ref{D25}) and (\ref{E7}).

The splitting (\ref{F5}) defines a bi-invariant flat connection $\nabla$
on the $\wC^{\times}$-bundle $\sG_{(-1)} @>>p> G_{(-1)}$.

\begin{pro}
The image of the monodromy map for $\nabla$,
\begin{equation*}
\Mon_{\nabla}\colon\pi_1\left(G_{(-1)},\Id\right)\to\wC^{\times}\,\,,
\end{equation*}
is a finite cyclic group consisting of roots of unity of order $2^m$,
where $0\le m\le[n/2]^2$, $n:=\dim M$.
\label{PF6}
\end{pro}

The origin of Proposition~\ref{PF6} lies in the K-theory. Namely,
we can take the direct sum of $E$ with another vector bundle $E_1$
such that $E\oplus E_1$ is a trivial vector bundle on $M$.

The group $G_{(-1)}^0$ (corresponding to zero order operators)
for the trivial $N$-dimensio-nal bundle ${\bold 1}_N$ on $M$ is
homotopy equivalent to the space of continuous maps
$$
P^*M\to GL(N,\wC)\sim U(N)\,\,,
$$
where $P^*M:=S^*M/\{\pm 1\}$.
By the Bott periodicity the fundamental group of the space $\Map(P^*M,U(N))$
stabilizes in the limit $N\to+\infty$ to $K^0(P^*M)$.

\begin{lem}
The monodromy $\Mon_{\nabla}$ is trivial for loops $\exp(2\pi it\sigma(p))$,
$0\le t\le 1$, in $G_{(-1)}^0$, where $p$ is a projector in $\End E$,
$p^2=p$ (considered as a zero order differential operator).
\label{LF8}
\end{lem}

The proof of Lemma~\ref{LF8} is based on analytic facts proved below
in this Section. (See Remark~\ref{RF2121}.)

Hence $\Mon_{\nabla}$ is defined by a homomorphism
\begin{equation}
\mon_{\nabla}\colon K^0(P^*M)/\pi^*K^0(M)\to\wC^{\times}\,\,,
\label{H5}
\end{equation}
$\pi\colon P^*M\to M$ is the natural projection map.

The orders of elements of the group $K^0(P^*M)/\pi^*K^0(M)$ are divisors
of $2^{([n/2]^2)}$ by the Atiyah-Hirzebruch spectral sequence
and by the fact that
$\widetilde{K}^0\left(\wR P^{n-1}\right):=K^0\left(\wR P^{n-1}\right)/\pi^*K^0
(pt)$ is the cyclic group $\wZ/2^{[n/2]}\wZ$ (see \cite{KV}, Lemma~4.5).

\subsection{Absolute value and holomorphic determinants}

For any invertible elliptic operator $A\in\Ell_{(-1)}^k$,
$k\in\wZ\setminus 0$, its absolute value determinant is defined by
\begin{equation}
|\det|A=\left({\det}_\zeta(A^*A)\right)^{1/2}\,\,.
\label{F16}
\end{equation}
Here, $A^*$ is the adjoint to $A$ operator with respect to a Hermitian
structure on $E$ and to a positive smooth density on $M$. The square root
on the right in (\ref{F16}) is positive. (The determinant
$\det_\zeta(A^*A)$ of a positive self-adjoint operator $A^*A$ is
 the usual one, i.e., it is taken with respect the spectral cut $\wR_-$.)

\begin{pro}
1. $|\det|A$, (\ref{F16}), is independent of a Hermitian structure and
of a positive density.

2. $|\det|(AB)=|\det|A\cdot|\det|B$ for $A,B\in\Ell_{(-1)}^{\wZ_+,\times}$.
Hence by multiplicativity we can extend $|\det|$ to a homomorphism
from $\Ell_{(-1)}^{\times}$ to $\wR_+^{\times}$.

3. The function $(|\det|A)^2$ on $\Ell_{(-1)}^{\times}$ can be locally
presented as $|f(A)|^2$, where $f$ is holomorphic in $A$.
\label{PF17}
\end{pro}

The assertions 1. and 2. are consequences of Theorem~\ref{TE9} and
of Lemma~\ref{LF14}. The assertion 3. is obtained by consideration
of a holomorphic function $(A,B)\to\det_\zeta(AB)$ for pairs $(A,B)$
close to $(A,A^*)$. By the multiplicativity property, Theorem~\ref{TE9},
(\ref{E10}), this function possesses the property
\begin{equation*}
{\det}_\zeta\left(A_1B_1\right){\det}_\zeta\left(A_2B_2\right)={\det}_\zeta
\left(A_1B_2\right){\det}_\zeta\left(A_2B_1\right)\,\,,
\end{equation*}
i.e., the matrix $\left(f_{A,B}\right)$, $f_{A,B}:={\det}_\zeta(AB)$,
has rank one. Hence locally $f_{A,B}=f_1(A)f_2(B)$, where $f_1$ and
$f_2$ are holomorphic. The restriction of $f_{A,B}$ to the diagonal
$\left\{(A,A^*)\right\}$ is real-valued. Multiplying $f_1$, $f_2$
by appropriate positive constants we obtain the assertion 3..

\begin{rem}
The assertion of Proposition~\ref{PF17}, {\em 1.}, in the case
of elliptic differential operators (of positive orders) was obtained
in \cite{Sch}. This result was one of the origins of the present
subsection.
\label{RH8}
\end{rem}

\begin{lem}
1. For $A\in H$ we have
$$
|\det|A=\left|{\det}_{Fr}(A)\right|\,\,.
$$

2. For $A\in\Aut E$ we have $|\det|A=1\,$.
\label{LF20}
\end{lem}

We use here $|\det|$ defined by multiplicativity (Proposition~\ref{PF17},
{\em 2.}) for operators from $\Ell_{(-1)}^0$.

\begin{rem}
By this lemma and by Proposition~\ref{PF17}
there is a Hermitian metric $\left\|\cdot\right\|_{\det}$
with {\em zero curvature} on the line bundle $L$ over $G_{(-1)}$
associated with the $\wC^{\times}$-bundle $\sG_{(-1)}$ over $G_{(-1)}$
such that
\begin{equation*}
|\det|A=\left\|d_1(A)\right\|_{\det}\,\,\,.
\end{equation*}
\label{RF21}
\end{rem}

\begin{pro}
The holomorphic flat connection $\nabla_{|\det|}$ on $L$ associated
with the metric $\left\|\cdot\right\|_{\det}$ coincides
with the connection $\nabla$ defined before Proposition~\ref{PF6}.
\label{PF22}
\end{pro}

One of possible proofs of this assertion follows from the invariance
of the connections $\nabla_{|\det|}$ and $\nabla$ with respect
to the natural action on $L$ of the group $\Diff(M,E)$.%
\footnote{Another proof of Proposition~\ref{PF22} is contained
in \cite{KV}, Proposition 6.12 (p.~108 of the preprint).}
This group is introduced in the proof of Lemma~\ref{LF14}.
Namely, any bi-invariant flat connection $\nabla_1$
on the $\wC^{\times}$-bundle $\sG_{(-1)} @>>p> G_{(-1)}$ defines
a homomorphism of Lie algebras
$a\left(\nabla_1\right)\colon\sfrg_{(-1)}\to\wC$ such that
$a\left(\nabla_1\right)|_{\wC}=\Id$ (where
$\wC\underset{j}{\hookrightarrow}\sfrg_{(-1)}$, (\ref{D20}), is the central
Lie subalgebra), and vice versa.

Thus $a\left(\nabla_{|\det|}\right)-a\left(\nabla\right)$ defines
a homomorphism
\begin{equation*}
b\colon\frg_{(-1)}\approx CS_{(-1)}^0\to\wC\,\,\,.
\end{equation*}
This homomorphism vanishes on $CS_{(-1)}^{-(n+1)}$, $n:=\dim M$.
Let us proof by induction (in $k$) that there are no nonzero continuous
linear functionals on $CS_{(-1)}^{-k}/CS_{(-1)}^{-(k+1)}$ invariant
with respect to the natural action of $\Diff(M,E)$. The step of this
induction is the assertion that the space of invariant continuous
linear functionals
$$
\left(\left(CS_{(-1)}^{-k}/CS_{(-1)}^{-(k+1)}\right)_{cont}^*\right)
^{\Diff(M,E)}
$$
is zero.

The quotient space $CS_{(-1)}^{-k}/CS_{(-1)}^{-(k+1)}$ is naturally
isomorphic to the space of $\wR^{\times}$-homogeneous
functions (with the values in $\End E$) on $T^*M\backslash M$
of order $-k$. The space of continuous linear functionals on it
is isomorphic via pairing by the noncommutative residue
$\left(,\right)_{\res}$ to the space of $\wR_+^{\times}$ homogeneous
generalized
functions $f$ of order $k-n$ such that $f(x,-\xi)=(-1)^{k-n+1}f(x,\xi)$.
(Here, $(-1)^{k-n+1}=(-1)^k$ as $n$ is odd.) But there is only one
(up to a constant factor) $\Diff(M,E)$-invariant generalized function
 with the values in $\End E$ on $T^*M\backslash M$. It is the identity
operator $\Id$. However, it corresponds to $k=n$
and has no appropriate central reflection symmetry.
Hence there are no nonzero $\Diff(M,E)$-invariant continuous linear
functionals on $CS_{(-1)}^0/CS_{(-1)}^{-(n+1)}$. Thus $b=0$ and
$\nabla=\nabla_{|\det|}$.

\begin{rem}
Let $L_1$ be the pullback of $L$ to the complex subgroup $\Aut E$
of elliptic operators of zero order. Then $L_1$ is trivialized
by the section
$d_1$ and the norm of this section with respect to the pullback
of $\left\|\cdot\right\|_{\det}$ is
identically equal to $1$ by Lemma~\ref{LF20}. Hence the monodromy of $L_1$
with respect to $\nabla$ is trivial by Proposition~\ref{PF22}.
This is equivalent to the assertion of Lemma~\ref{LF8}.
\label{RF2121}
\end{rem}

\subsection{Trace type functional for odd class operators}

For $A\in CL_{(-1)}^{\wZ}$ define $\TR_{(-1)}(A)$ as the value at zero
of the function $\zeta_{A,C}(s)=\TR\left(AC^{-s}\right)$, where
$C\in\Ell_{(-1)}^{2k}$, $k\in\wZ_+$, is sufficiently close
to a positive definite self-adjoint one. The correctness of the definition
above is the content of the following lemma.

\begin{lem}
1. $\zeta_{A,C}(s)$ is holomorphic at $s=0$.

2. $\zeta_{A,C}(0)$ is independent of $C$.
\label{LF201}
\end{lem}

The assertion 1. follows from the equalities
$$
\Res_{s=0}\left(\zeta_{A,C}(s)\right)={1\over 2k}\res(A)=0\,\,\,.
$$
(These equalities hold by Theorem~\ref{TB3} and by Lemma~\ref{LE3}.)
The assertion 2. is a consequence of Theorem~\ref{TB3}, (\ref{A25}),
applied to a holomorphic at $s=0$ family
$\left(AC^{s/\ord C}-AB^{s/\ord B}\right)/s$. Namely
\begin{equation}
\left(\zeta_{A,C}(s)-\zeta_{A,B}(s)\right)|_{s=0}=\left(\sigma(A),{\log
_{(\theta)}\sigma(C)\over\ord C}-{\log_{(\theta)}\sigma(B)\over\ord B}\right)
_{\res}.
\label{F202}
\end{equation}
Here, $L_{(\theta)}$ are admissible spectral cuts close to $\wR_-$.
The difference of the logarithms in (\ref{F202}) belongs to $CS_{(-1)}^0$
by Proposition~\ref{PE6}, (\ref{E7}). Thus $\zeta_{A,C}(0)=\zeta_{A,B}(0)$.

\begin{lem}
1. The linear functional $\TR_{(-1)}$ coincides with the usual trace $\Tr$
on $CL_{(-1)}^{-(n+1)}$. Hence it is equal to zero
on $\frh^{(1)}\subset CL_{(-1)}^{-(n+1)}$.

2. The induced by $\TR_{(-1)}$ linear functional on $\sfrg_{(-1)}$
is equal to the splitting homomorphism $a(\nabla)\colon\sfrg_{(-1)}\to\wC$
 defined in the proof of Proposition~\ref{PF22}, see also (\ref{F5}).

3. For $A,B\in CL_{(-1)}^{\wZ}$ we have
\begin{equation}
\TR_{(-1)}([A,B])=0\,\,\,,
\label{F204}
\end{equation}
i.e., $\TR_{(-1)}$ is a trace type functional on $CL_{(-1)}^{\wZ}$.
\label{LF203}
\end{lem}

The first assertion follows from Theorem~\ref{TB3}, {\em 1.} (i.e.,
from the equality $\TR=\Tr$ on trace class operators). The proof
of the second assertion is the same as the proof of Proposition~\ref{PF22}.
(In that proof we use only the invariance under the natural action
of $\Diff(M,E)$.)

The trace property (\ref{F204}) in the case $A,B\in CL_{(-1)}^0$ follows
from the assertion 2. as $a(\nabla)$ is a Lie algebra homomorphism.
(We don't give here a proof for $A$ or $B$ of positive orders
and don't use in this text the property (\ref{F204}) for general $A$, $B$.)

For an operator $C\in CL_{(-1)}^0$ we can define an entire function
\begin{equation}
\zeta_C^{\TR_{(-1)}}(s):=\TR_{(-1)}(\exp(-sC))\,\,\,.
\label{F205}
\end{equation}
(Here we use Lemma~\ref{LF10}.) The determinant
of $\exp C\in\Ell_{(-1)}^0$ corresponding to this zeta-function
is equal to
\begin{equation}
{\det}_{(-1)}(\exp C,C):=\exp\left(-\df_s\zeta_C^{\TR_{(-1)}}(s)|_{s=0}\right)=
\exp\left(\TR_{(-1)}C\right)\,\,.
\label{F206}
\end{equation}
The last equality in (\ref{F206}) follows from Theorem~\ref{TB3}, {\em 3.},
applied to a holomorphic in two variables family $\exp(-zC)\cdot B^{-s}$
 where $B\in\Ell_{(-1)}^2$ is positive definite. Namely, by this
theorem $\TR\left(\exp(-zC)B^{-s}\right)$ is holomorphic in $(s,z)$
for $0<|s|<1/2$. Also by this thorem and by Lemma~\ref{LE3}, (\ref{E4}),
there are no singularities at $s=0$. So
\begin{equation*}
\df_z\TR_{(-1)}(\exp(-zC))|_{z=0}=\left(\df_z\TR\left(\exp(-zC)B^{-s}\right)
|_{z=0}\right)\big|_{s=0}=-\TR_{(-1)}C\,\,\,.
\end{equation*}

\begin{rem}
The following statement holds . Let for $A\in\Ell_{(-1)}^{2k}$, $k\ne 0$,
its principal symbol $\sigma_{2k}(A)$ possess a spectral cut $\theta$
(i.e., let the Agmon-Nirenberg condition hold for $A$). Then for $m\in\wZ$
\begin{equation*}
\zeta_{A,\theta}(-m)=\TR_{(-1)}\left(A^m\right)\,\,.
\end{equation*}
\label{RF221}
\end{rem}

Let $A\in\Ell_{(-1)}^0$ and let $\log\sigma(A)=:x\in CS_{(-1)}^0$ exist.
Then a canonical determinant of $A$ is defined by
\begin{equation}
{\det}_{(-1)}(A,x)=d_1(A)d_{0,(-1)}(\sigma(A),x)^{-1}\,\,,
\label{F222}
\end{equation}
where
$$
d_{0,(-1)}(\sigma(A),x):=d_1(\exp X)\exp\left(-\TR_{(-1)}X\right)
$$
for any $X\in CL_{(-1)}^0$ with $\sigma(X)=x$. The proof of independence
of $d_{0,(-1)}(\sigma(A),x)$ of a choice of $X$ (with given $x$) uses
the equality
$$
d_1\left(\exp X\exp\left(-X_1\right)\right)={\det}_{Fr}\left(\exp X\exp\left(
-X_1\right)\right)
$$
for $\sigma(X)=\sigma\left(X_1\right)=x$, and
the assertion for $\det_{(-1)}$ analogous to (\ref{A20}), Lemma~\ref{L1}.
(See \cite{KV}, Lemma~6.12.)

\begin{rem}
1. The projective microlocal Agmon-Nirenberg condition (Section~\ref{S5.1})
is sufficient for the existence of $\log\sigma(A)$ in $CS_{(-1)}^0$.

2. An element $d_{0,(-1)}(\sigma(A),x)$ (in (\ref{F222})) can be defined
by any homotopy
class of paths in $G_{(-1)}$ from $\Id$ to $\sigma(A)$, namely,
using the flat connection $\nabla$ on the $\wC^{\times}$-bundle
$\sG_{(-1)} @>>p> G_{(-1)}$.
\label{RF223}
\end{rem}

\begin{lem}
The determinant $\det_{(-1)}(A,x)$ for an operator $A\in\Ell_{(-1)}^0$
close to positive definite ones and for $x$ defined by a spectral cut
$\wR_-$ is equal
to the determinant $\det(A)$ introduced in (\ref{E15}) with the help
of the multiplicative property (\ref{E10}).
\label{LF224}
\end{lem}

This assertion follows from Lemma~\ref{LF203}.
\section{Open problems}

1. To construct the determinant group $\sG$ globally in terms
of symbols (i.e., in particular, without using analytic continuations
and the Fredholm determinants).

This problem is not solved even for the connected component $\sG_c$
of $\Id\in\sG$. By Theorem~\ref{TD22} we know a description
of the Lie algebra of $\sG_c$ in terms of symbols. Thus we can
reconstruct the universal cover $\wG$ of $\sG_c$,
\begin{equation*}
\sG_c\simeq\wG/\Gamma\,\,\,,
\end{equation*}
where $\Gamma$ is a discrete subgroup of the center $Z(\wG)$.
We have some information about $\Gamma$ because
$\sG_c/\wC^{\times}=\SEll_0$ is the connected component of $\Id$
of the group of elliptic symbols.
It is enough to have a description of the restrictions
of the $\wC^{\times}$-central extension $\sG_c @>>p> G_c$
to one-parameter compact subgroups $T_q:=\exp(2\pi itq)$, where
$q\in CS^0=\frg_0\subset\frg\,\,$ is a projector in symbols, $q^2=q$,
and $0\le t\le 1$.

These subgroups are generators of the fundamental group
$\pi_1\left(G_c,\Id\right)$. The preimage $\sT_q:=p^{-1}\left(T_q\right)$
of $T_q$ in $\sG_c$ is a $3$-dimensional
abelian Lie group,
\begin{equation*}
1\to\wC^{\times}\to\sT_q @>>p> S^1\left(=T_q\right)\to 1\,\,\,.
\end{equation*}
To describe this central extension, we choose an element
$q_1$ of the Lie algebra
$\operatorname{Lie}\left(\!\sT_q\!\right)\!\subset\sfrg$
of $\sT_q$, $q_1=q\mod\wC$. Then the nonzero complex number
$$
\exp\left(2\pi iq_1\right)=:c=c\left(q_1\right)\in\wC^{\times}\subset\sT_q
$$
defines $\sT_q$ because of the identification
\begin{equation*}
\sT_q\simeq\wC^{\times}\times\wR/(c,1)\cdot\wZ\,\,\,.
\end{equation*}
To define such an element $q_1$, it is enough to choose
$x\in\frg=S_{\log}$ with $\ord x=1$ and set $q_1:=\Pi_{\sx}q$.
(See Proposition~\ref{PD3}, {\em 1.}, and the proof of Theorem~\ref{TD22}.)

The number $c\left(\Pi_{\sx}q\right)$ can be expressed in terms
of a spectral invariant of a pair $(P,X)$ of a PDO-projector $P$
with $\sigma(P)=q$ (such a projector always exists) and of an element
$X\in\fell$ with $\sigma(X)=x$. Namely
\begin{equation}
c\left(\Pi_{\sx}q\right)=\exp(-2\pi if(P,X))\,\,\,,
\label{H4}
\end{equation}
where $f(P,X)=\TR(P\exp(sX))|_{s=0}(\mod\wZ)$. (Note that
by the equality (\ref{H4}) the element $f(P,X)\in\wC/\wZ$ is
independent of choices of $P$ and $X$.) The equality (\ref{H4})
is proved in \cite{KV}, Proposition~7.1. Hence $f(P,X)$ is a function
of $q=\sigma(P)$, $x=\sigma(X)$. We call it the {\em generalized
spectral asymmetry}. If $\exp X$ is self-adjoint and $P$ is the orthogonal
projector to the linear subspace spanned by the eigenvectors of $\exp X$
with positive eigenvalues, this invariant is simply expressed via
the spectral asymmetry of $\exp X$, \cite{APS}. However $f(P,X)$
cannot be obtained as a value (taken modulo $\wZ$) of an integral
of a local in symbols $\sigma(P)$, $\sigma(X)$ expression. Thus
the description of $\sG_c$ in terms of symbols reduces to computation
of the generalized spectral asymmetry in terms of symbols (and without
using of liftings to PDOs). \\

2. To generalize constructions and results of this paper to the case
of elliptic complexes. \\

3. To compute the homomorphism (\ref{H5}) for odd-dimensional
manifolds. (This is the monodromy of the holomorphic determinant
for odd class operators.)

Estimates of Proposition~\ref{PF6} for the torsion
of $\Image\left(\mon_{\nabla}\right)$ are probably not the best possible.
They were produced by bounding the orders for the elements
in $K^0(P^*M)/\pi^*K^0(M)$. \\

4. To investigate analytic properties of entire functions
$\zeta_C^{\TR_{(-1)}}(s)$ for $C\in CL_{(-1)}^0$ defined in (\ref{F205}).

Do these functions have representations in terms of Dirichlet series?

\end{document}